\documentclass[aps,showpacs,twocolumn,prb]{revtex4-1}
\usepackage{graphicx}
\usepackage{amsmath,amssymb,amsfonts}
\usepackage{color}

\begin{document}

\title{Anisotropy Reversal of the Upper Critical Field at Low Temperatures and Spin-Locked Superconductivity in K$_{2}$Cr$_{3}$As$_{3}$}

\author{F. F. Balakirev$^1$}
\author{T. Kong$^2$}
\author{M. Jaime$^1$, R. D. McDonald$^1$, C. H. Mielke$^1$}
\author{A. Gurevich$^3$}
\author{P. C. Canfield$^2$} 
\author{S. L. Bud'ko$^2$}
\affiliation{$^1$National High Magnetic Field Laboratory, Los Alamos National Laboratory, MS-E536, Los Alamos, New Mexico 87545, USA}
\affiliation{$^2$Ames Laboratory, US DOE and Department of Physics and Astronomy, Iowa State University, Ames, Iowa 50011, USA}
\affiliation{$^3$Department of Physics, Old Dominion University, Norfolk, VA 23529, USA}

\begin{abstract}
We report the first measurements of the anisotropic upper critical field $H_{c2}(T)$ for  K$_{2}$Cr$_{3}$As$_{3}$ single crystals up to 60 T and $T > 0.6$ K. Our results show that the upper critical field  parallel to the Cr chains, $H_{c2}^\parallel (T)$, exhibits a paramagnetically-limited behavior, whereas the shape of the $H_{c2}^\perp (T)$ curve (perpendicular to the Cr chains) has no evidence of paramagnetic effects. As a result, the curves $H_{c2}^\perp (T)$ and $H_{c2}^\parallel(T)$ cross at $T\approx 4$ K, so that the anisotropy parameter $\gamma_H(T)=H_{c2}^\perp/H_{c2}^\parallel (T)$ increases from $\gamma_H(T_c)\approx 0.35$ near $T_c$ to $\gamma_H(0)\approx 1.7$ at 0.6 K. This behavior of $H_{c2}^\|(T)$ is inconsistent with triplet superconductivity but suggests a form of singlet superconductivity with the electron spins locked onto the direction of Cr chains. 
 
\end{abstract}


\maketitle

Recently, superconductivity in K$_{2}$Cr$_{3}$As$_{3}$ with the transition temperature $T_{c}$ = 6.1 K\cite{Bao15} was discovered, followed by the reports on related superconducting compounds Rb$_2$Cr$_3$As$_3$ with $T_c=4.8$ K \cite{Tang}, and Cs$_2$Cr$_3$As$_3$ with $T_c = 2.2$ K \cite{TangCs}.  These materials have attracted considerable attention because their crystalline lattices contain  an array of weakly coupled, double well [(Cr$_3$As$_3)^{2-}]_{\infty}$ linkages stretched along the $c$-axis, suggesting the possibility of quasi-one dimensional (1D) superconductivity in inorganic materials. Specific heat and transport measurements on both polycrystals and single crystals \cite{Bao15,Kong15a} have shown that $T_c$ is not very sensitive to the sample quality quantified by the residual resistance ratio (RRR).  Both NMR and penetration depth measurements on K$_{2}$Cr$_{3}$As$_{3}$ suggest strong, quasi-1D paramagnetic fluctuations and unconventional superconductivity with possible line nodes in the order parameter\cite{Zhi,Pang15}.  Density functional theory (DFT) calculations show that hybridization of $3d$-orbitals of Cr and $4p$ orbitals of As results in two quasi-1D $\alpha$ and $\beta$ bands and a 3D $\gamma$ band crossing the Fermi surface, but no spontaneous magnetic order \cite{el1,el2}.  

Measurements of the upper critical field $H_{c2}(T)$ at $H \leq 14$ T performed on K$_{2}$Cr$_{3}$As$_{3}$ single crystals \cite{Kong15a} revealed moderately anisotropic $H_{c2}(T)$ with very large initial slopes, $dH_{c2}^\|/dT=12$ T/K along the Cr chains and $dH_{c2}^\perp/dT=7$ T/K  perpendicular to the chains. Extrapolation of these $H_{c2}(T)$ curves to $T=0$ suggests the values of $H_{c2}(0)$ two-three times larger than the BCS paramagnetic limit, $H_p$ [T]$=1.84T_c$ [K]. Similarly large initial $H_{c2}(T)$ slope observed in the low-field measurements on a polycrystalline  Rb$_2$Cr$_3$As$_3$ was interpreted as a hint of possible triplet superconductivity \cite{Tang}. In  K$_{2}$Cr$_{3}$As$_{3}$, the anisotropy of $H_{c2}(T)$ measured at $H < 14$ T diminishes as $T$ decreases so crossing of the $H_{c2}(T)$ curves appears likely \cite{Kong15a}. Given the ambiguity of conclusions based on the extrapolations of $H_{c2}(T)$ measured near $T_c$ to low temperatures, we performed high-field measurements of $H_{c2}(T)$ on K$_{2}$Cr$_{3}$As$_{3}$ single crystals in pulsed magnetic fields up to $H=60$ T which enabled us to reveal the full anisotropic $H_{c2}(T)$ curves from $T_c$ down to 600 mK. The results turned out to be striking: whereas $H_{c2}^\parallel(T)$ parallel to the Cr chain exhibits a clear Pauli-limited behavior, the perpendicular $H_{c2}^{\perp}(T)$ shows no sign of paramagnetic pairbreaking and becomes significantly larger than $H_{c2}^\parallel(T)$ at $T\ll T_c$. We show that this behavior of $H_{c2}(T)$ is inconsistent with triplet superconductivity and propose an alternative interpretation which also addresses the apparent contradiction of the moderate anisotropy of $H_{c2}$ with the putative 1D superconductivity in K$_{2}$Cr$_{3}$As$_{3}$.    

K$_{2}$Cr$_{3}$As$_{3}$ single crystals were grown using a high-temperature solution growth method\cite{Canfield92,Kong15a}. The starting stoichiometry was K:Cr:As = 6:1:7 following Ref.~\onlinecite{Bao15}. Bulk, elemental K, Cr and As were held in an alumina crucible that was welded in a tantalum tube under argon atmosphere. A third layer of silica tube protects tantalum from getting oxidized at high temperature. The whole ampoule was then slowly heated up to 1000 $^{\circ}$C and cooled down over 3 days to 700 $^{\circ}$C, at which temperature the remaining liquid was separated from the crystals in a centrifuge. Detailed experimental set-up and temperature profile can be found in Ref.\onlinecite{Kong15a}. Single crystals of K$_{2}$Cr$_{3}$As$_{3}$ are rod-like and very sensitive to air, so care was taken to prevent oxidation of samples. The Cr chains in the crystal structure are parallel to the rod.

\begin{figure}[tbh!]
\includegraphics[scale = 0.3]{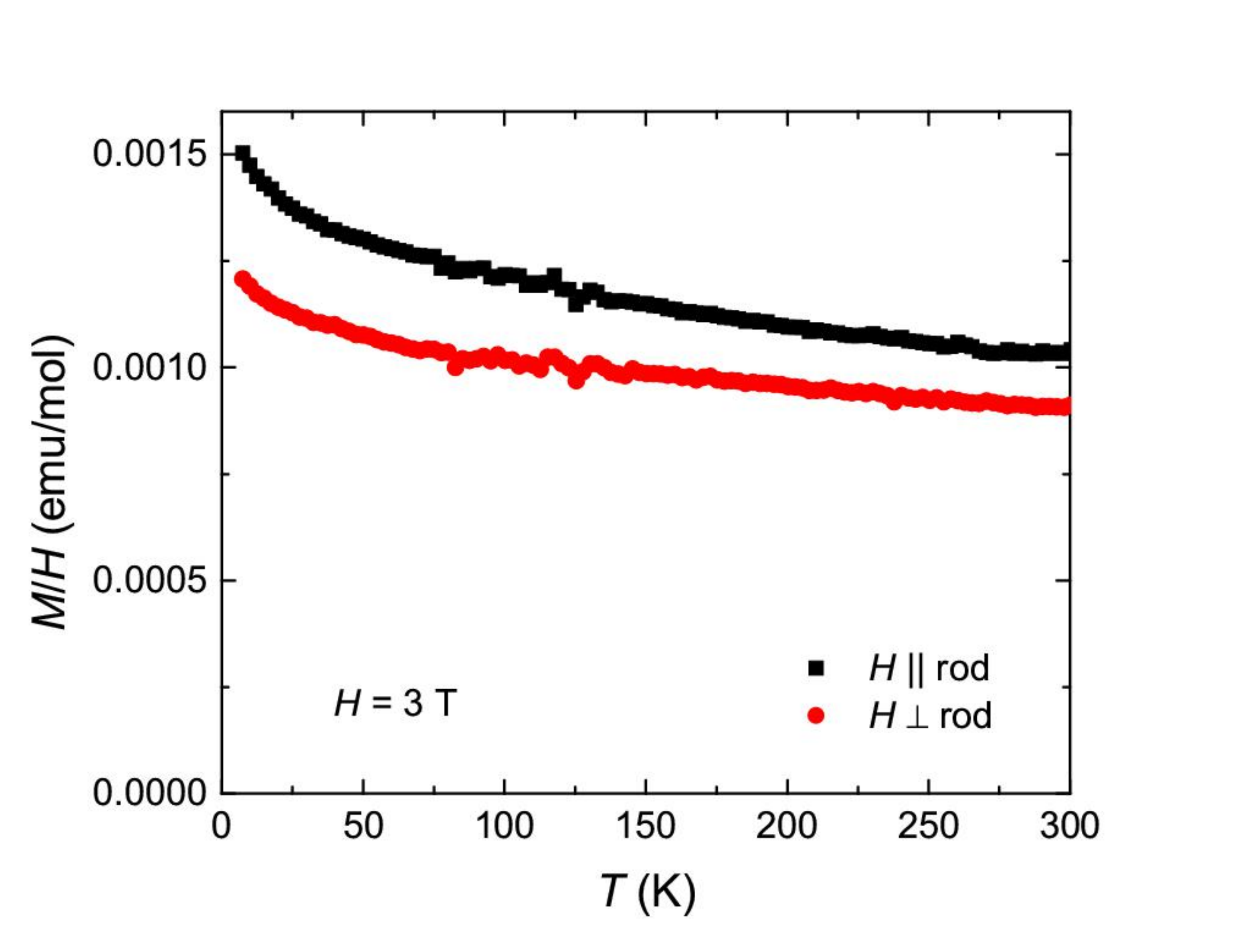}
\caption{(Color online) Anisotropic dc magnetic susceptibility of K$_{2}$Cr$_{3}$As$_{3}$ measured at 3 T for $H \parallel$ rod (black squares) and $H \perp$ rod (red circles).}
\label{ani}
\end{figure}

The anisotropic dc susceptibility $\chi(T) = M(T)/H$ of the samples in the normal state was measured using a Quantum Design Magnetic Property Measurement System (MPMS). A Kel-F\textsuperscript{\textregistered} plastic disk with a small amount of Apiezon\textsuperscript{\textregistered} N grease was used to hold several aligned samples. The diamagnetic background signal was then subtracted from the raw data to obtain the sample signal. Temperature dependencies of $\chi_\|(T)$ parallel to the Cr chains and $\chi_\perp(T)$ perpendicular to the chains measured at $7<T<300$ K are shown in Fig.~\ref{ani}. Both $\chi_\|(T)$ and $\chi_\perp(T)$ increase as $T$ decreases, with $\chi_\|(T)\simeq 1.2\chi_\perp(T)$ over the whole temperature range.

\begin{figure}[tbh!]
\includegraphics[scale = 0.5]{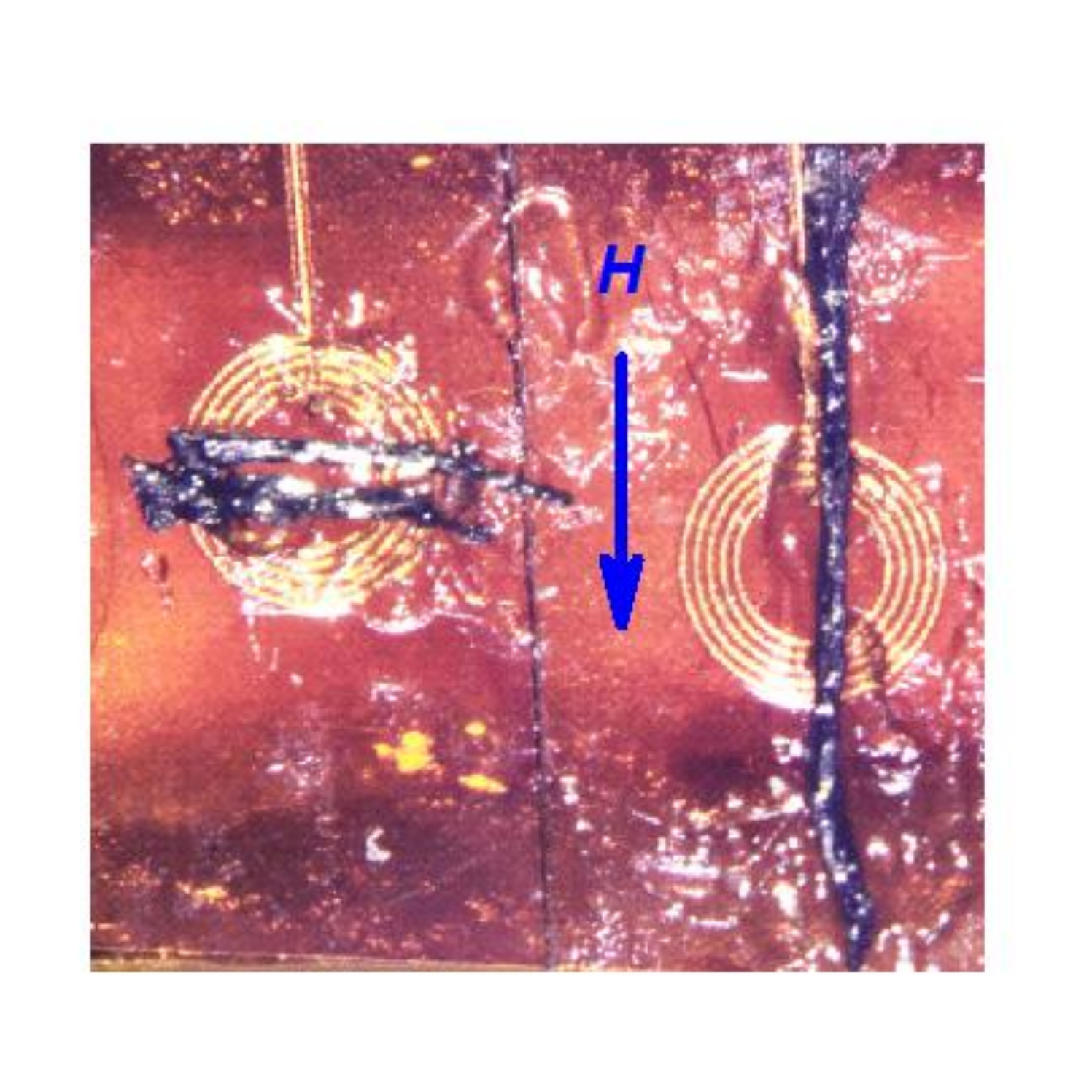}
\caption{(Color online) Parts of a single crystal sample mounted on resonating radio frequency (RF) coils. The coils are lithographically defined planar spiral coils 0.8 mm in diameter, coil geometry preferable for coupling to rod-like samples. Magnetic field direction is shown by the blue arrow.}
\label{resexp}
\end{figure}

The upper critical field $H_{c2}$ below 4 K was measured in a 65 T pulsed magnet at the National High Magnetic Field Laboratory (NHMFL), Los Alamos, using a contactless technique based on a proximity detector oscillator (PDO) \cite{alt09a,mun11a}. A K$_{2}$Cr$_{3}$As$_{3}$ crystal was separated into several pieces and placed, one parallel to field and the other perpendicular to field, on two separate spiral copper coils (see Fig.~\ref{resexp}). The lithographically defined resonating RF coils were designed at NHMFL. The coils are connected to a PDO resonating in the 30-35 MHz range. The oscillatory signal waveforms were recorded directly using a fast 200 million samples per second digitizer during 120 ms magnet pulse. 

\begin{figure}[tbh!]
\includegraphics[scale = 0.3]{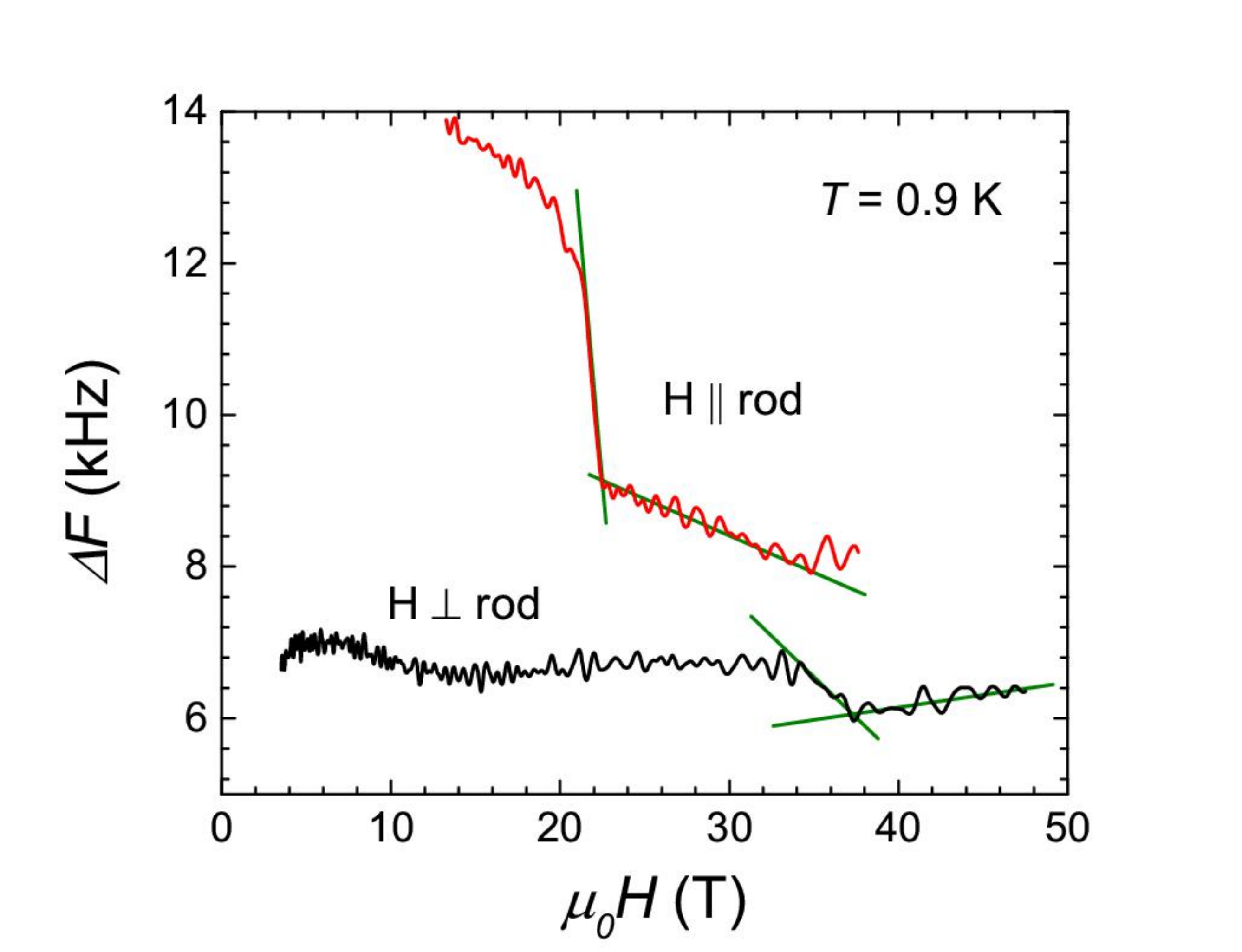}
\caption{(Color online) Magnetic field dependence of the resonance frequency in PDO circuits for samples oriented along the field and perpendicular to the field at $T = 0.9$ K after subtraction of background signal. The background signal is a polynomial fit to frequency response recorded at $T$ = 8 K where no superconductivity is present. Green lines indicate the criterion for determining $H_{c2}$ (see text).}
\label{res}
\end{figure}

A smooth featureless PDO frequency response to magnetic field up to 60 T was observed above $T_c$ at 8 K. In contrast, at $T<T_c$ there is a noticeable change in the resonance frequency dependence as a function of field. As the magnetic flux is partially expelled by the superconductor, the effective inductance of the coils is reduced and the resonance frequency increases. Once the magnetic field exceeds $H_{c2}$, the normal state frequency response is recovered again. The onset of superconductive response,\em{ i.e. }\rm first deviation from the high-field normal state behavior, in RF oscillator and magnetotransport measurements has been shown to track the value of $H_{c2}$ obtained by other experimental probes \cite{MilekeHC2,AltarawnehHC2,TailleferHC2}. We extracted the intercept point of two types of behavior and associate the so-obtained value of the magnetic field with $H_{c2}$ as shown in Fig.~\ref{res} for a representative  temperature.

\begin{figure}[tbh!]
\includegraphics[scale = 0.3]{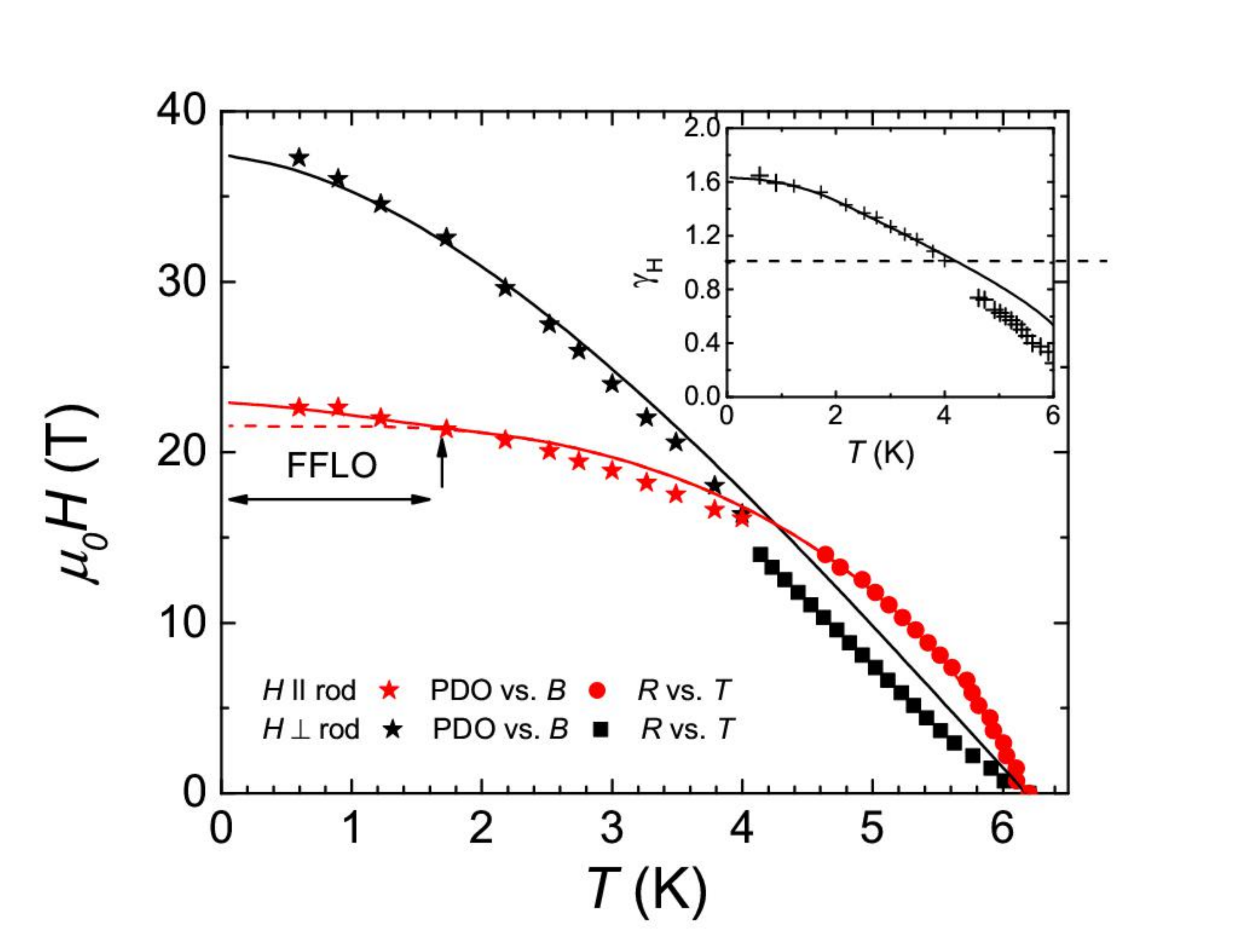}
\caption{(Color online) Upper critical fields $H_{c2}^{\parallel}(T)$ and $H_{c2}^{\perp}(T)$.  Data below 14 T were obtained using a superconductivity onset criterion  (as in Ref.~\onlinecite{nin08a}) on resistance vs temperature data taken at constant field in Ref.~\onlinecite{Kong15a}. Data below 4 K were obtained in $^{3}$He liquid below 1 K and in $^{4}$He liquid above 1 K in pulsed fields using PDO onset criteria. Solid lines are fits of $H_{c2}^\|(T)$ to Eq. (\ref{U}) for $\alpha=2.6$ and $H_{c2}^\perp(T)$ for $\alpha=0$.  For $\textbf{H}||c$, the fit suggests a FFLO state below 1.8 K. Inset shows $\gamma_H(T)=H_{c2}^{\perp}/H_{c2}^\|$ where the solid line was calculated from the fits. }
\label{Hc2}
\end{figure}

Shown in Fig.~\ref{Hc2} are combined $H_{c2}(T)$ data for K$_{2}$Cr$_{3}$As$_{3}$ obtained from resistivity measurements \cite{Kong15a} below 14 T and PDO measurements in pulsed magnetic field  down to 600 mK. For the resistivity data, the onset of resistive transition (as opposed to offset in Ref.~\onlinecite{Kong15a}) was used, which is more consistent with the PDO criterion outlined above. The apparent mismatch between two criteria/two sets of data is small, $< 0.05$ K (or $< 0.5$ T). The temperature dependencies of $H_{c2}(T)$ along two crystal orientations show distinctly different behaviors. For field applied along the $c$-axis, $H_{c2}^{\parallel}(T)$ exhibits a strong negative curvature and saturation at $\sim$ 23 T. By contrast, $H_{c2}^{\perp}(T)$ perpendicular to the $c$-axis is roughly linear in $T$ with a slight upward curvature near $T_c$ and equally sight negative curvature at lower $T$ where $H_{c2}^{\perp}$ reaches $\sim$ 37 T.  The $H_{c2}^{\parallel}(T)$ and $H_{c2}^{\perp}(T)$ curves cross at $T\approx 4$ K, so that the anisotropy, parameter $\gamma(T)=H_{c2}^{\perp}/H_{c2}^{\parallel}$ increases as $T$ decreases, from $\gamma_H\approx 0.35$ at $T_c$ to $\gamma_H(T)>1$ below $T\approx 4$ K. Crossing of $H_{c2}(T)$ curves has also been observed on quasi 1D organic superconductors\cite{Lee97,org} and some of the Fe-based superconductors \cite{Fang10,Maiorov14}. 

Essential features of the $H_{c2}(T)$ curves shown in Fig.~\ref{Hc2} set them apart from other materials mentioned above. The values of $H_{c2}^{\parallel}(0)\approx 2 H_p$ and $H_{c2}^\perp(0)\simeq 3.7 H_p$ are both well above the BCS paramagnetic limit, $H_p$[T]$= 1.84 T_c [K] \simeq 11$ T for $T_c=6$ K. This fact could be interpreted as evidence of triplet superconductivity \cite{Tang}, but this would be inconsistent with other observations: (1) $T_c$ is practically insensitive to nonmagnetic impurities as evident from the measurements of $T_c$ on samples with different RRR's \cite{Bao15,Kong15a}, (2) The shape of $H_{c2}^{\parallel}(T)$ shown in Fig.~\ref{Hc2} is indicative of strong Pauli pairbreaking, similar to what was observed in Fe-based superconductors \cite{ag,Kogan14,tar,bnl}. Thus, the behavior of $H_{c2}^\|(T)$ is characteristic of singlet superconductivity, yet the shape of $H_{c2}^{\perp}(T)$ indicates orbital pairbreaking with no apparent Pauli-limiting effects \cite{whh}. The latter is rather remarkable, given that in a quasi-1D compound it is the transverse $H_{c2}^{\perp}(T)$ which is expected to be limited by both orbital and Pauli pairbreaking so that $H_{c2}^{\perp}(T)<H_{c2}^{\parallel}(T)$. Moreover, the $H_{c2}(T)$ curves shown in Fig.~\ref{Hc2} do not exhibit strong upturns at $T\ll T_c$ indicative of  superconductivity coexisting with field-induced spin density waves in organic materials \cite{os1,os2}, Fulde-Ferrel-Larkin-Ovchinnikov (FFLO) states \cite{fflo} or orbital multiband effects \cite{ag}. The inversion of anisotropy of $H_{c2}(T)$ was also associated with equatorial nodes in a single band order parameter  \cite{Kogan14}, but this model disregards the essential Pauli-limiting effects. High $H_{c2}$ values well above $H_p$ but no crossover of the $H_{c2}(T)$ curves were observed in other quasi-1D inorganic superconductors Li$_{0.9}$Mo$_6$O$_{17}$ (Ref. \onlinecite{Mercure}),  Nb$_2$Pd$_x$Se$_5$ (Ref. \onlinecite{Khim}), and Nb$_2$Pd$_{0.81}$S$_5$ (Ref. \onlinecite{Zhang}). The high values of $H_{c2}$ observed on these materials, as well as $H_{c2}^\perp$ shown in Fig. 4 can hardly be explained by the enhancement of $H_{c2}$ by the spin orbital effects \cite{bnl,whh} which would require unrealistically large spin-orbital coupling constants \cite{Mercure,Khim}.

We suggest that the behavior of $H_{c2}(T)$ shown in Fig. 4 can result from a particular spin configuration of pairing electons, given that strong exchange correlations in K$_{2}$Cr$_{3}$As$_{3}$ can result in local energy minima for spins either parallel or perpendicular to the Cr chains in the normal state, as was shown by DFT calculations \cite{el1,el2}.  The high-field measurements of $H_{c2}$ could thus probe the spin orientation of the itinerant electrons in bands responsible for superconductivity if the Zeeman energy shift $ \mu_BH <2.5$ meV at $H<40$ T does not cause spin flip transitions between different states separated by energy barriers $>2.5$ meV. In this case three possible spin configuration of singlet Cooper pairs can manifest themselves in markedly different behaviors of $H_{c2}(T)$:      

1. The direction of the antiparallel spins in Cooper pairs is not fixed, thus both $H_{c2}^\|(T)$ and $H_{c2}^\perp(T)$ would be paramagnetically-limited, since $H_{c2}^\perp(0) > 3H_p$. Because $H_{c2}^\perp(T)$ is further reduced by orbital pairbreaking, there should be no crossing of $H_{c2}^\perp (T)$ and $H_{c2}^\|(T)$. This conventional scenario is inconsistent with our data.

 2. The Cooper pair spins are predominantly aligned perpendicular to the Cr chains, so that spins in each octahedral cluster of Cr chains can either all point inward or alternate antiferromagnetically \cite{el1,el2}.  In that case $H_{c2}^\|$ would be unaffected by the Pauli pairbreaking, while $H_{c2}^\perp$ would be Pauli limited, which is opposite to what is shown in Fig. 4.
 
3. The Cooper pair spins are predominantly aligned along the Cr chains, which thus play the role of an easy magnetization axis but cause no magnetic order in the paramagnetic normal state. This scenario is consistent with the observed Pauli-limiting behavior of $H_{c2}^{\parallel}(T)$ and no signs of it in $H_{c2}^{\perp}(T)$. Such spin alignment is consistent with the NMR measurements\cite{Zhi} which revealed strong 1D spin correlations in K$_2$Cr$_3$As$_3$, and also with the sign of the anisotropy of $\chi(T)$ shown in Fig. 1, although the uniaxial anisotropy of $\chi(T)$ in the normal state of non-local-moment-bearing materials is not unusual \cite{bud99a,kon14a}.     

K$_2$Cr$_3$As$_3$ is likely a multiband superconductor \cite{el1,el2}, but currently little is known about the pairing mechanism and symmetry of the order parameter. To get insight into the behavior of $H_{c2}(T)$ without inrtoducing too many unknown parameters, we first use a  Werthamer-Helfand-Hohenberg theory \cite{whh} for a uniaxial, single band superconductor. The equation for $H_{c2}$ which takes into account anisotropic orbital pairbreaking in the clean limit and the Pauli effects for the spins aligned along the $c$-axis, but disregards spin-orbital effects is given by  \cite{ag}.
    \begin{gather}
    \ln t + 2e^{q^{2}}\operatorname{Re}\sum_{n=0}^{\infty}\int_{q}^{\infty}due^{-u^{2}}\times \nonumber \\ \left[\frac{u}{n+1/2}-\frac{t}{\sqrt{b}}\tan^{-1}\left(
    \frac{u\sqrt{b}}{t(n+1/2)+i\alpha b}\right)\right]=0,
    \label{U} \\
    b=\frac{\hbar^{2}v_\perp^{2}\epsilon_{\theta}^{1/2}H_{c2}}{8\pi\phi_{0}
    T_{c}^{2}g^2},\qquad\alpha=\frac{4\mu_B\phi_{0}T_{c}g}{\hbar^{2}v_\perp^2\epsilon_{\theta}^{1/2}}\cos\theta,
    \label{ba} \\
    \quad q^{2}=\frac{Q_{z}^{2}\epsilon\phi_{0}}{2\pi H\epsilon_{\theta}^{3/2}},\qquad \epsilon_\theta = (\cos^2\theta+\epsilon\sin^2\theta)^{1/2}.
    \label{q}
    \end{gather}
Here $t=T/T_c$, $v_\perp$ is the Fermi velocity in the $ab$ plane, $\epsilon = m_\perp/m_\|\gg 1$, $m_\|$ is the band mass of electrons moving along the chains and $m_\perp$ is the effective mass for hopping across the chains, $\phi_0$ is the magnetic flux quantum, $\mu_B$ is the Bohr magneton, $\theta$ is the angle between $\textbf{H}$ and the $c$-axis, $g=1+\lambda$ describes strong-coupling Eliashberg corrections \cite{carb}, $\lambda$ is a dimensionless pairing constant, and $\alpha$ is the Pauli pairbreaking parameter. For $\alpha\gtrsim 1$, the transition to the FFLO state with a spatially modulated order parameter occurs \cite{fflo}. For the field tilted away from the $c$-axis, the factor $\cos\theta$ in $\alpha$ takes into account the projection of $\textbf{H}$ onto the spin direction along the chains, so that no Pauli pairbreaking occurs at $\theta=\pi/2$.

Shown in Fig. 4 is the fit of Eq. (\ref{U}) to the experimental data using $H_{c2}(0)$ and $\alpha$ as the only fit parameters for each of the $H_{c2}^\|(T)$ and $H_{c2}^\perp(T)$ curves. Here $H_{c2}^{\parallel}(T)$ is described well by Eq. (\ref{U}) with $\theta=0$ and $\alpha=2.6$. As follows from Fig. 4, the FFLO vector  $Q(T)$ should appear spontaneously below 1.8 K where the solid $H_{c2}^\parallel (T)$ curve lies above the dashed curve calculated for $Q=0$. The value of $H_{c2}^\parallel(0)\approx 23$ T at $T=0$ exceeds the weak-coupling BCS paramagnetic limit but is consistent with the paramagnetic limit enhanced by strong-coupling effects $\tilde{H}_p=(1+\lambda )H_p$ (Ref. \onlinecite{carb}). The observed $\tilde{H}_p\simeq H_{c2}^\perp(0) = 23$ T corresponds to $\lambda\approx 1.1$ indicative of strong pairing correlations in K$_2$Cr$_3$As$_3$ (Ref. \onlinecite{Bao15}).    

If spins are locked onto the Cr chains, $H_{c2}^\perp(T)$ is only limited by orbital pairbreaking. The fit is satisfactory although deviations near $T_c$ and at $T\ll T_c$ are apparent. Such deviations are generally expected as the orbitally-limited $H_{c2}^\perp (T)$ is more sensitive to multiband effects, symmetry of the order parameter, broadening of the resistive transition due to local $T_c$ inhomogeneities, and the different criteria for $H_{c2}$ in resistive and PDO measurements, than the paramagnetically-limited $H_{c2}^\|(T)$.  The fit of $H_{c2}^\perp(T)$ can be improved using a multiband version of Eq. (\ref{U})  (Ref. \onlinecite{ag}) and taking into account the actual band structure and possible nodes in the order parameter \cite{Kogan14}, but this would greatly increase the number of adjustable  parameters.

Analysis of the slopes $H_{c2}'=dH_{c2}/dT$ at $T_c$ can give addtional insights into features of superconductivity in K$_2$Cr$_3$As$_3$. The anisotropic Ginzburg-Landau (GL) theory gives $H_{c2}'^\parallel=\phi_0/2\pi\xi_\perp^2 T_c$ and $H_{c2}'^\perp=\phi_0/2\pi\xi_\perp\xi_\parallel T_c$, where $\xi_\perp(T)=\epsilon^{-1/2}\xi_\parallel(T)$ and $\xi_\|(T)$ are the GL coherence lengths across and along the chains, and $m_\perp/m_\|=[H_{c2}'^\parallel/H_{c2}'^\perp]^2$. The observed $H_{c2}'^\parallel\simeq 12$ T/K and $H_{c2}'^\perp\simeq 7$ T/K yield a modest mass anisotropy  $m_\perp/m_\|\approx 3$, hardly consistent with the putative 1D superconductivity in K$_2$Cr$_3$As$_3$. Indeed, $H_{c2}^\perp(0)\approx 37$ T implies that the transverse coherence length $\xi_\perp(0) = [\phi_0/2\pi\sqrt{\epsilon}H_{c2}^\perp(0)]^{1/2}\approx 2.2$ nm is more than twice the spacing between the Cr octahedrals in the $ab$ plane, thus  the in-plane Josephson coupling between the Cr chains is not negligible even at $T=0$. The reduced anisotropy of $H_{c2}(T)$ may also result from a contribution of the 3D $\gamma$ band, similar to MgB$_2$ where the isotropic $\pi$ band greatly reduces the anisotropy of $H_{c2}$ as compared to the prime 1-D $\sigma$ band \cite{mgb2}. For an  effective 1D $\alpha\beta$ band coupled with the 3D $\gamma$ band, $H_{c2}'$ can be evaluated using a two-band theory \cite{mgb2}
 \begin{equation}
\frac{dH_{c2}}{dT}=\frac{24\pi\phi_0k_B^2T_c}{7\zeta(3)\hbar^2(c_+v_1^2+c_-v_2^2)},
\label{tb}
\end{equation}   
where $v_1(\theta)$ and $v_2(\theta)$ are Fermi velocities in $\alpha\beta$ and $\gamma$ bands, 
$2c_\pm=1\pm\lambda_-/\lambda_0$, $\lambda_0=(\lambda_-^2+4\lambda_{12}\lambda_{21})$, $\lambda_-=\lambda_{11}-\lambda_{22}$, $\lambda_{11}$ and $\lambda_{22}$ are dimensionless pairing constants in bands 1 and 2, and $\lambda_{12}$ and $\lambda_{21}$ are interband pairing constants. The 3D band 2 can significantly reduce the anisotropy of $H_{2}$ even if superconductivity in band 2 is proximity induced by the main 1D bands, as it happens in MgB$_2$ (Ref. \onlinecite{mgb2}). If superconductivity in the $\gamma$ band is mostly induced by interband coupling, the Pauli pairbreaking effects would be reduced even if the spins in the $\gamma$ band are not locked onto the Cr chains.  

In conclusion, we report the first high-field measurements of the full temperature dependence of anisotropic $H_{c2}(T)$ in K$_2$Cr$_3$As$_3$. The temperature dependence of $H_{c2}^\parallel(T)$ parallel to the Cr chains exhibits a clear Pauli-limited behavior and possible FFLO state below 1.8 K. However, $H_{c2}^\perp (T)$ perpendicular to the chains shows an orbitally-limited behavior with no signs of Pauli pairbreaking. As a result, the curves $H_{c2}^\perp (T)$ and $H_{c2}^\parallel(T)$ cross at $T\approx 4$ K, so that the anisotropy parameter $\gamma_H(T)=H_{c2}^\perp/H_{c2}^\parallel (T)$ increases from $\gamma_H(T_c)\approx 0.35$ at $T_c$ to $\gamma_H(0)\approx 1.7$ at 0.7 K. Our results seem inconsistent with triplet superconductivity but could be interpreted in terms of singlet superconductivity with electron spins of 1D bands locked onto the Cr chains.  The modest anisotropy of $H_{c2}$ suggests in-plane Josephson coupling of the Cr chains and a contribution of a 3D $\gamma$ band. 
 
We thank V. Kogan and B. Ramshaw for useful discussions. Work done in Ames Laboratory is supported by the U.S. Department of Energy, Office of Science, Basic Energy Sciences, Materials Science and Engineering Division under contract NO. DE-AC02-07CH11358. The NHMFL Pulsed Field Facility is supported by the National Science Foundation, the Department of Energy, and the State of Florida through NSF cooperative grant DMR-1157490 and by US DOE BES  "Science at 100T" project.


\bibliographystyle{apsrev4-1}
%

\end{document}